\documentclass[twocolumn, tighten, twocolappendix]{aastex631}

\newcommand{\ebvstar}{$E(B-V)_{\rm star}$}
\newcommand{\ebvgas}{$E(B-V)_{\rm gas}$}
\newcommand{\hii}{H~{\sc ii}}
\newcommand{\haew}{$\rm {EW}_{H\alpha}$}
\newcommand{\havd}{$\rm \sigma_{H\alpha}$}
\newcommand{\logten}{$\log_{10}$}
\newcommand{\hasb}{$\rm \Sigma_{H\alpha}$}
\newcommand{\sigmamass}{$\Sigma_*$}
\newcommand{\hasbunit}{$\rm erg\;s^{-1}\;kpc^{-2}$}
\newcommand{\oiilr}{[O~{\sc ii}]$\;\lambda\lambda3726,3729$}
\newcommand{\oiiilr}{[O~{\sc iii}] $\lambda \lambda 4959, 5007$}
\newcommand{\oiiir}{[O~{\sc iii}]$\lambda 5007$}
\newcommand{\oiii}{[O~{\sc iii}]}
\newcommand{\niilr}{[N~{\sc ii}] $\lambda \lambda 6548, 6583$}
\newcommand{\niir}{[N~{\sc ii}]$\lambda 6583$}
\newcommand{\nii}{[N~{\sc ii}]}
\newcommand{\siilr}{[S~{\sc ii}] $\lambda \lambda 6717, 6731$}
\newcommand{\siil}{[S~{\sc ii}]$\lambda 6717$}
\newcommand{\sii}{[S~{\sc ii}]}
\newcommand{\niioii}{[N{\sc ii}]/[O{\sc ii}]}
\newcommand{\oiiioii}{[O{\sc iii}]/[O{\sc ii}]}
\newcommand{\niisii}{[N{\sc ii}]/[S{\sc ii}]}
\newcommand{\hiiexplorer}{{\tt HII{\scriptsize EXPLORER}}}

\defcitealias{Li_N_2020}{Paper I}
\defcitealias{Li_N_2021}{Paper II}

\shorttitle{dust attenuation scaling relations}
\shortauthors{Li et al.}
\begin{document}

\title{Estimating Dust Attenuation From Galactic Spectra. III. Radial variations of dust attenuation scaling relations in MaNGA galaxies}

\correspondingauthor{Niu Li \& Cheng Li}

\author[0000-0002-0656-075X]{Niu Li}
\affiliation{Department of Astronomy, Tsinghua University, Beijing 100084, China}
\email{liniu@tsinghua.edu.cn}

\author[0000-0002-8711-8970]{Cheng Li}
\affiliation{Department of Astronomy, Tsinghua University, Beijing 100084, China}
\email{cli2015@tsinghua.edu.cn}

\begin{abstract}
We investigate the radial dependence of the scaling relations of 
dust attenuation in nearby galaxies using integral field spectroscopy (IFS)
data from MaNGA. We identify ionized gas regions of kpc sizes 
from MaNGA galaxies, and for each region we estimate both the 
stellar attenuation \ebvstar\ and gas attenuation \ebvgas. We then
quantify the correlations of 15 regional/global properties with  
\ebvgas\ and \ebvstar, using both the feature importance 
obtained with the Random Forest regression technique and the 
Spearman correlation coefficients. The importance of 
stellar mass, metallicity and nebular velocity dispersion found 
previously from SDSS-based studies can be reproduced if our 
analysis is limited to the central region of galaxies. The scaling 
relations of both \ebvgas\ and \ebvstar\ are found to strongly vary 
as one goes from the galactic center to outer regions, and from 
H$\alpha$-bright regions to H$\alpha$-faint regions. For \ebvgas, \niisii\ 
is top ranked with a much higher correlation coefficient than any 
other property at $0<R\lesssim R_e$, while \oiiioii\ outperforms \niisii\ 
as the leading property in the outermost region. For \ebvstar, stellar 
age shows the strongest correlation with no/weak dependence on 
radial distance, although \hasb\ and sSFR present similarly strong 
correlations with \ebvstar\ in the galactic center. We find H$\alpha$-bright regions 
to generally show stronger correlations with \ebvgas, while H$\alpha$-faint 
regions are more strongly correlated with \ebvstar, although 
depending on individual properties and radial distance. The 
implications of our results on studies of high-$z$ galaxies are discussed. 
\end{abstract}

\section{Introduction} \label{sec:introduction}

Ubiquitously distributed in interstellar medium (ISM) and accounting for 
only $\sim$1\% of the mass of a typical galaxy
\citep[e.g.][]{Remy-Ruyer-2014,Driver-2018}, dust forms through
the ejecta of asymptotic giant branch (AGB) stars and supernovae 
\citep[e.g.][]{Dwek_E_1998, Popping_G_2017}, grows by 
accreting gas-phase metals \citep[][]{Dominik_C_1997,Dwek_E_1998,
Zhukovska_S_2014}, and can be destroyed through supernova shocks,
thermal sputtering and grain collisions, or incorporated into newly 
formed stars \citep[e.g.,][]{Dwek_E_1998,Bianchi_S_2005,Nozawa_T_2007}.
Interstellar dust is composed of grains of different sizes ranging 
from 5 to 250 nm typically \citep{Weingartner2001}, and made of 
various chemical elements \citep{Draine_BT_2003ARA&A}.
The size, mass and chemical composition of dust grains may evolve 
through various mechanisms \citep{Asano2013}. 

Dust plays important roles in  the thermodynamics and chemistry of 
the ISM as well as star formation and radiative transfer in the host galaxy 
\citep[e.g.][]{Conroy_C_2013ARA&A}. Dust can facilitate 
star formation in galaxies by acting as a catalyst in the formation 
of molecular Hydrogen \citep{Draine_BT_2003ARA&A} and cooling
down the ISM through thermal emission in infrared \citep{Montier2004}
and depletion of metals onto dust grains \citep{Schneider2006}.
Dust grains may absorb or scatter the light emitted by stars, thus
reshaping the spectral energy distribution (SED) of the galaxy, an 
effect known as dust attenuation or dust extinction 
\citep{Galliano_F_2018ARA&A,Salim_S_2020ARA&A}. 
Dust attenuation laws can be generally characterized by three distinct
properties: the dust attenuation curve (attenuation as a function of 
wavelength) which is approximately a power law in optical
and UV, the UV bump which accounts for the excess attenuation
at around 2175{\AA}, and the relation between the stellar and nebular
attenuation which is usually quantified by comparing the optical color
excess \ebvstar\ and \ebvgas\ as measured from the stellar continuum 
and the Balmer emission lines. In a recent review article, \citet{Salim_S_2020ARA&A} 
summarized previous studies of the attenuation curve and 
UV bump in galaxies of different redshifts, which have been mostly 
limited to global measurements of the galaxies. Spatially-resolved 
dust attenuation curves and the UV bump strength have recently been 
derived for a sample of nearby galaxies based on the {\it Swift}/UVOT data
\citep[e.g.][]{Molina2020b,Belles2023,Duffy2023,Zhou2023}.

The correlation between \ebvgas\ and \ebvstar\ was originally studied 
by \cite{1988ApJ...334..665F} who found \ebvgas\ to be significantly 
higher than \ebvstar. \citet{Calzetti1994,Calzetti_D_2000} analyzed a
sample of local starburst galaxies and found a typical ratio of the 
stellar-to-gas attenuation of $f\equiv$\ebvstar/\ebvgas$\approx0.44$.
Extensive studies in the past two decades in different types of 
galaxies and at different redshifts have confirmed the larger attenuation 
in gas than in stellar populations, but finding $f$ to vary over a wide 
range, from 0.44 to $\sim1$ 
\citep[e.g.,][]{Calzetti_D_2000, Riffel_R_2008, Reddy_NA_2010,
Wuyts_S_2011, Kashino_D_2013, Kreckel_K_2013, Wuyts_S_2013,
Price_SH_2014,Pannella_M_2015, Valentino_F_2015, Puglisi_A_2016,
Zahid_HJ_2017, Buat_V_2018, Koyama_Y_2019, Qin_JB_2019}. 
The discrepancy between \ebvgas\ and \ebvstar\ may be generally 
explained by a two-component dust model in which the dust disk
of a galaxy is formed of two components: a diffuse and optically-thin 
component distributed throughout the ISM, plus dense and 
optically-thick clouds (birth clouds) where young stars are born
\citep[e.g.,][]{Charlot_S_2000, Wild_V_2011, Chevallard_J_2013}.
In this model, the emission lines 
produced in star-forming regions are attenuated by both the dust 
in the birth clouds and that in the ambient ISM, while the continuum 
radiation of older stars is attenuated only by the diffuse dust in ISM. 
Consequently, if spatially-resolved observations are available, one 
would expect a stronger correlation between \ebvgas\ and \ebvstar\ 
in star-forming regions than in regions dominated by diffuse ionized gas (DIG),
as well as a strong dependence of the value of $f$ on the 
average age of stars in the ionized regions. In fact, recent studies 
of the spatially-resolved \ebvgas\ and \ebvstar\ in low-$z$ galaxies 
based on integral field spectroscopy data from the Mapping Nearby 
Galaxies at Apache Point Observatory \citep[MaNGA;][]{Bundy_K_2015}
survey have produced consistent results with this model 
\citep[e.g.,][]{Greener2020,Lin_ZS_2020,Li_N_2021,Riffel_R_2021}.

There have been many studies focused on the gas attenuation only,
which can be easily measured from the Balmer decrement  
(H$\alpha$/H$\beta$) for galaxies at different redshifts, aiming to 
understand how \ebvgas\ depends on the physical properties of 
galaxies. These studies have established that, among many 
properties and followed by star formation rate and gas-phase metallicity, 
the stellar mass appear to be the most important property in driving 
\ebvgas\ for both low-redshift galaxies \citep[e.g.,][]{Asari_NV_2007,
Garn_T_2010,Maheson_G_2023} and high-redshift galaxies \citep[e.g.,][]{Pannella_M_2009,Garn2010b,Whitaker2017,McLure2018,
Shapley_AE_2022,Shapley_AE_2023}. As pointed out by \cite{Maheson_G_2023},
the strong dependence of dust attenuation on stellar mass is expected
from simple analytical relations between dust-to-gas mass, stellar mass
and metallicity. However, the previous studies have been largely based 
on single-fiber or slitless spectroscopy, thus limited to dust attenuation 
and galaxy properties measured in the central region of low-$z$ galaxies, 
or the global measurements for galaxies at higher redshifts. The use 
of IFS data should provide valuable insights into the scaling relations of 
dust attenuation which essentially rely on how the two dust components 
are distributed across the galaxy. Based on the MaNGA data,
\citet{Lin_ZS_2020} and \citet[][hereafter Paper II]{Li_N_2021} 
have recently highlighted the significance of gas-phase metallicity 
and ionization level in relation to \ebvgas\ at kpc scales, with the 
\niisii\ flux ratio being particularly identified as the most important property.
The discrepancy between the IFS-based studies and those based on
single-fiber spectroscopy data should be partly (if not fully) attributed 
to the different spatial resolutions of the data used in different studies.
The discrepancy also strongly implies that the scaling relations of dust 
attenuation vary from region to region within galaxies. 

This work is the third of a series of papers dedicated to spatially-resolved 
dust attenuation in MaNGA galaxies. In \citet[][hereafter Paper I]{Li_N_2020},
we developed a new technique to estimate a model-independent 
attenuation curve from the observed optical spectrum for each spaxel 
in MaNGA datacubes. In \citetalias{Li_N_2021}, we applied this technique 
to approximately 8000 unique galaxies from MaNGA to investigate the 
correlation between \ebvgas\ and \ebvstar\ in both DIG 
and \hii\ regions, examining how the correlation depends on the physical 
properties of galaxies. A total number of 16 regional or global properties 
were considered. In this work, we extend the study of \citetalias{Li_N_2021}
by further examining how the scaling relations of both \ebvgas\ and \ebvstar\ 
depend on the radial distance of the ionized regions. We will firstly 
concentrate on the central region of our galaxies in order to make 
comparisons with previous studies of single-fiber spectroscopy data. 
We then consider the regions at all different radial distances. We use 
Random Forest regression to assess the feature importance of the 
galactic properties, and also use the Spearman correlation coefficient
to quantify the correlation strength between dust attenuation and the 
properties. 

The paper is organized as follows. In \autoref{sec:data} we describe the 
observational data, the quantities measured from the data, and the 
samples for our analysis. We then present our results in \autoref{sec:result}
and discuss on the results in \autoref{sec:discussion}. Finally, we summarize in
\autoref{sec:summary}. Throughout this paper we assume a $\Lambda$ 
cold dark matter cosmology model with $\Omega_{\rm m}=0.3$, 
$\Omega_\Lambda=0.7$, and $H_0=70 \;{\rm km\;s^{-1}\;Mpc^{-1}}$. 
We assume a \cite{Chabrier_G_2003} initial mass function (IMF).

\section{Data} \label{sec:data}

The data used in this work is the same as in \citetalias{Li_N_2021}. 
In this section we briefly describe the data and measurements, 
and refer the reader to \S2 of \citetalias{Li_N_2021} for a comprehensive
description.

\subsection{MaNGA}

As one of the key components of the fourth-generation Sloan
Digital Sky Survey project \citep[SDSS-IV;][]{Blanton_MR_2017},
the MaNGA survey obtained IFS data for 10,010 nearby galaxies
over its six-year operation from July 2014 through August 2020
 \citep[][]{Bundy_K_2015}.  MaNGA utilize 17 hexagonal 
science integral field units with a field of views ranging from 
12$^{\prime\prime}$ to 32$^{\prime\prime}$ to obtain IFS datacubes
covering out to 1.5 or 2.5 effective radii ($R_e$) of the target galaxies
\citep{Drory_N_2015}. The target galaxies are selected from the 
SDSS-based catalog, NASA Sloan Atlas \citep[NSA][]{Blanton_MR_2011} 
so as to cover a stellar mass range of 
$5\times10^8M_\odot h^{-2}\leq M_\ast \leq 3\times10^{11}M_\odot h^{-2}$ 
and a redshift range of $0.01<z<0.15$ with a median redshift of 
$z\sim0.03$ \citep{Wake_DA_2017}. The MaNGA spectra span a 
wavelength range from 3622 \AA\ to 10354 \AA\ with a spectral resolution
$R\sim2000$ and an $r$-band signal-to-noise (S/N) ratio of 4$-$8 
per \AA\ per 2\arcsec\ fiber at $1-2R_e$ of the galaxies. 

MaNGA raw data are reduced with the Data Reduction Pipeline
\citep[DRP;][]{Law_DR_2016} which produces a datacube for each 
target galaxy with a spaxel (spatial pixel) size of 0\arcsec.5$\times$0\arcsec.5
and an effective spatial resolution of $\sim$2\arcsec.5. The absolute 
flux calibration of the MaNGA spectra is $<5\%$ for $>80\%$ of 
the wavelength range. Spectral flux calibration, survey execution 
strategy and data quality tests are detailed in \citet{Yan_RB_2016a} 
and \citet{Yan_RB_2016b}. Additionally, the Data Analysis Pipeline (DAP) 
performs full spectral fitting to the reduced datacubes, deriving 
measurements of stellar kinematics, emission lines, and spectral 
indices \citep{Westfall_KB_2019,Belfiore2019}. The final data  
from MaNGA including DRP and DAP products of 11,273 datacubes 
for 10,010 unique galaxies are released as a part of the SDSS Data 
Release 17 \citep[SDSS DR17;][]{Abdurro_uf_2022}.
In this work, we make use of an earlier sample, the MaNGA Product 
Launch 9, which includes 8113 datacubes for 8000 unique
galaxies, a random subset of the final sample in SDSS/DR17.

\subsection{Identifying Ionized Gas Regions}\label{subsec:data_igas}

We first identify ionized gas regions in the MaNGA galaxies, and 
for each region we then quantify the stellar and gas attenuation, along 
with other relevant properties. Following the methodology described 
in \cite{Liang_FH_2020}, we apply the publicly available \hii-region finder
\hiiexplorer\ \citep{Sanchez_SF_2012} to the H$\alpha$ flux map 
to detect the ionized gas regions. Originally intended for identifying \hii\
regions, \hiiexplorer\ commonly applies a relatively high threshold
density, for instance, H$\alpha$ surface brightness \hasb$>10^{39.5}$ \hasbunit.
Instead, we opt for a much lower threshold of \hasb$>10^{37.5}$ \hasbunit\ 
to include DIG regions in our analysis. 
The identified ionized gas regions each consist of a few to tens of spaxels
with similar H$\alpha$ surface densities by definition. We stack the 
spectra within each region to derive an averaged spectrum with a 
high S/N \citep{Liang_FH_2020}, from which we then calculate the 
stellar and gas attenuation, as well as other pertinent properties required 
for our study.

Due to the limitation in resolution, it is not possible to resolve 
individual \hii\ regions, which typically range in size from a few to 
hundreds of parsecs \citep[e.g.,][]{Kennicutt_RC_1984,Hunt_LK_2009,
Anderson_LD_2019}. As a result, each of the ionized gas regions identified 
from MaNGA, which have a size of approximately 1 kpc, may encompass 
a few to hundreds of individual \hii\ regions, or represent a combination 
of DIG and \hii\ regions. \cite{Zhang_K_2017} found that \hasb\ can be 
effectively utilized to differentiate DIG-dominated regions from \hii-dominated 
regions within the MaNGA galaxies, with an empirical dividing value of 
\hasb$\sim10^{39}$ \hasbunit. Accordingly we divide the ionized regions 
in our sample into two subsets, and for simplicity we will refer to 
\hii-dominated regions as H$\alpha$-bright regions and DIG-dominated regions 
as H$\alpha$-faint regions in what follows.

\subsection{Measuring gas and stellar attenuations and regional properties}

For each of the ionized regions as identified above, we firstly obtain a 
model-independent attenuation curve ($A_\lambda-A_V$) by applying 
the technique developed in \citetalias{Li_N_2020} to the stacked spectrum 
of the region. 
In this method, the observed spectrum is first decomposited into two components,
separately including the small-scale features ($S$) and large-scale spectral shape ($L$),
by applying a moving-average filter. The spectrum of each single stellar-population 
model template is also separated into the $S$ and $L$ components in the same way. 
The intrinsic dust-free model spectrum of the stellar component is then obtained 
by fitting the observed ratio of the small- to large-scale spectral components 
$(S/L)_{\rm obs}$ to the corresponding ratio of the model spectra $(S/L)_{\rm model}$. 
Since the small- and large-scale components are affected by dust 
in the same way, their ratio $S/L$ is expected to be dust-free, as long as the 
dust attenuation curves are similar for different stellar populations 
in a galactic region. Finally, $A_\lambda-A_V$ is determined by comparing 
the observed spectrum with the best-fit model spectrum.
Subsequently, the value of \ebvstar, defined as $A_B-A_V$, 
is obtained from the attenuation curve. 

Next, we use the attenuation curve to correct the effect of 
dust attenuation for the observed spectrum. We then perform full spectral fitting 
to the dust-free spectrum, and measure stellar population parameters from 
the best-fit spectrum of the stellar component as well as emission line 
parameters from the starlight-subtracted spectrum. For the spectral 
fitting, we employ the simple stellar population (SSP) library provided 
by \citet[hereafter BC03]{Bruzual_Charlot_2003}, which provides spectra 
for a comprehensive set of 1326 SSPs with 221 different ages ranging 
from $t=0$ Gyr to $t=20$ Gyr and six metallicities from $Z=0.005 Z_\sun$ 
to $Z = 2.5 Z_\sun$. The computation of SSPs is based on the
Padova evolutionary track \citep{Bertelli_G_1994} and adopts the
stellar initial mass function (IMF) of \cite{Chabrier_G_2003}. We select a 
subset of 150 SSPs, covering 25 ages that are evenly spaced in \logten$t$ 
at each of the six metallicities. The effect of stellar velocity 
and velocity dispersion is taken into account by shifting/broadening the 	
SSP model spectra according to the stellar velocity $v_\ast$ and velocity 
dispersion $\sigma_\ast$ measurements provided by the MaNGA DAP.
Based on the best-fit stellar spectrum, we then obtain the following parameters 
to quantify the stellar populations in each region: surface density of 
stellar mass $\Sigma_\ast$, light-weighted stellar age $t_L$,
light-weighted stellar metallicity $Z_L$, and the narrow-band version
of the 4000\AA\ break $D_n4000$ defined by \cite{Balogh_ML_1999}.

We subtract the best-fit stellar spectrum from the stacked spectrum, and
we use Gaussian profiles to fit different emission lines, thus measuring the 
flux, surface density, velocity dispersion, and equivalent width (EW) of 
the emission lines within each region. The lines considered particularly 
in this work include \oiilr, H$\beta$, \oiiilr, \niilr, H$\alpha$, and \siilr.
The fluxes have been corrected for attenuation, using the 
H$\alpha$-to-H$\beta$ flux ratio from the observed spectrum and assuming 
the case-B recombination:
\begin{equation}
    E(B-V)_{\rm gas} = \frac{2.5}{k(\lambda_{\rm H\beta}) - k(\lambda_{\rm H\alpha})}
    \log_{10}\left[\frac{(\rm H\alpha/\rm H\beta)
    _{\rm obs}}{2.86}\right],
    \label{eqn:def_ebvbd}
\end{equation}
where $k(\lambda_{\rm H\beta})$ and $k(\lambda_{\rm H\alpha})$  
are the attenuation at the wavelength of H$\alpha$ and H$\beta$
given by the same attenuation curve as obtained above for the stellar component.
Derived from the measurements of emission lines, we have estimated the
following parameters to characterize properties related to the gas
component:
H$\alpha$ surface density \hasb, H$\alpha$ equivalent width \haew,
H$\alpha$ velocity dispersion \havd,
logarithm of specific star formation rate (SFR) defined by the ratio of
SFR to stellar mass sSFR, gas-phase metallicity 12+\logten(O/H)
estimated from the parameter O3N2$\;\equiv\;$(\oiiir/H$\beta$)/(\niir/H$\alpha$)
\citep{Pettini_M_2004},
logarithm of the flux ratio \logten(\niioii) between \niir\ and \oiilr,
logarithm of the flux ratio \logten(\oiiioii) between \oiiir\ and \oiilr,
logarithm of the flux ratio \logten(\niisii) between \niir\ and \siil.
In the subsequent sections, we employ shorthand notations in certain figures
to enhance visual clarity: $Z_{gas}$=12+\logten(O/H), N2O2=\logten\niioii,
O3O2=\logten\oiiioii, N2S2=\logten\niisii.

In our analysis, we refine our focus to a subset of ionized gas regions
exhibiting significantly elevated S/N in both the stellar continuum and
the relevant emission lines. Specifically, we stipulate a requirement of
S/N$>5$ for both the stellar continuum and the H$\alpha$ and H$\beta$
lines, while a S/N$>3$ criterion is imposed for the \oiilr, \oiiir,
\niir, and \siilr\ lines. Additionally, we exclude regions classified
as AGN on the Baldwin-Phillips-Terlevich diagram \citep[BPT;][]{Baldwin_JA_1981}.
We first exclude all the regions that are classified as Seyferts in the
\oiii/H$\beta$ versus \sii/H$\alpha$ diagram. We also exclude the regions
that are located within 3\arcsec\ from their galactic center and are
classified as either low-ionization emission-line regions (LINER) 
in the \oiii/H$\beta$ versus \sii/H$\alpha$ diagram
or as active galactic nuclei (AGN) in the \oiii/H$\beta$ versus \nii/H$\alpha$ diagram.
The final sample includes $\sim 2.7\times10^5$ ionized gas regions, of which 
$\sim40\%$ are H$\alpha$-faint regions and $\sim60\%$ are H$\alpha$-bright regions.

\subsection{Regional and global properties in consideration}

This work is aimed to investigate the dependency of both \ebvgas\ 
and \ebvstar\ on a variety of physical properties at various radial distances, 
both at regional and global scales. In alignment with \citetalias{Li_N_2021}, 
we consider 12 regional properties: 
\logten\hasb, \logten\haew, sSFR, \logten\sigmamass,
\logten$t_L$, $Z_L$, $D_n4000$, 12+\logten(O/H),
\logten\niioii, \logten\oiiioii, \logten\niisii, \logten\havd.
Note that when compared to \citetalias{Li_N_2021}, we have omitted 
$t_M$ and $Z_M$ due to their similarities with $t_L$ and $Z_L$, while 
additionally including \havd\ which was found in \cite{Maheson_G_2023} 
to be strongly correlated with \ebvgas. 
Furthermore, we consider three global properties: total stellar 
mass ($M_\ast$), morphological type ($T$-type), and the $r$-band 
minor-to-major axial ratio ($b/a$). The $M_\ast$ and $b/a$ measurements 
are derived from NSA, and the $T$-type parameters are taken from 
\cite{Dominguez_S_2018}. A negative value of $T$-type typically corresponds
to early-type morphology, whereas a positive $T$-type value signifies
late-type morphology. In total, we consider a total of 15 properties
including 12 regional properties and 3 global properties. 

\begin{figure*}
    \fig{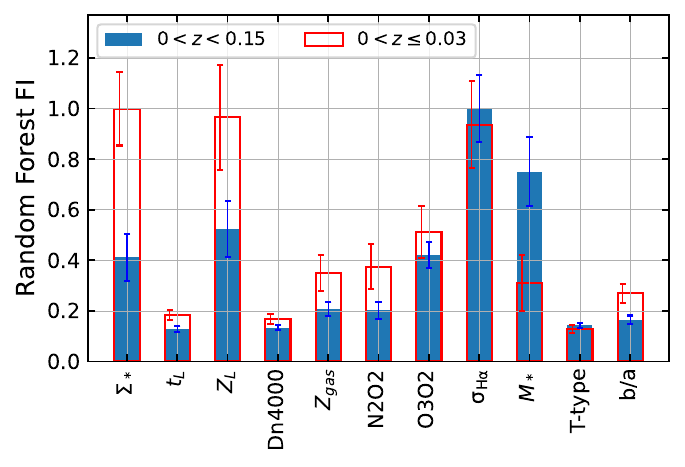}{0.48\textwidth}{(a)}
    \fig{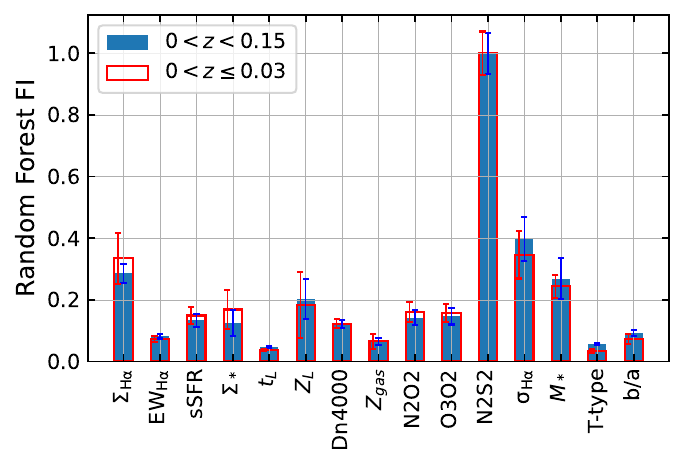}{0.48\textwidth}{(b)}
    \caption{Random Forest regression is employed to analyze the 
    relationship between \ebvgas\ and galactic properties in the central region 
    of galaxies.
    The histograms display the FI of the properties.
    Note that the maximum FI has been normalized to 1.
    Panel (a) excludes properties associated with H$\alpha$ flux and \niisii,
    whereas Panel (b) encompasses the entirety of the 15 properties
    under consideration. In each panel the solid/blue histogram and red histogram
    show the results respectively for the full sample and the subsample at $z\leq 0.03$. 
	The errorbar shows standard deviation of  feature importance, estimated by bootstrap 
	resampling of all the spaxels in the sample and re-training the Random Forest 
	regression for 100 times.
	}
    \label{fig:rf_center}
\end{figure*}

\begin{figure*}
	\epsscale{1.15}
	\plotone{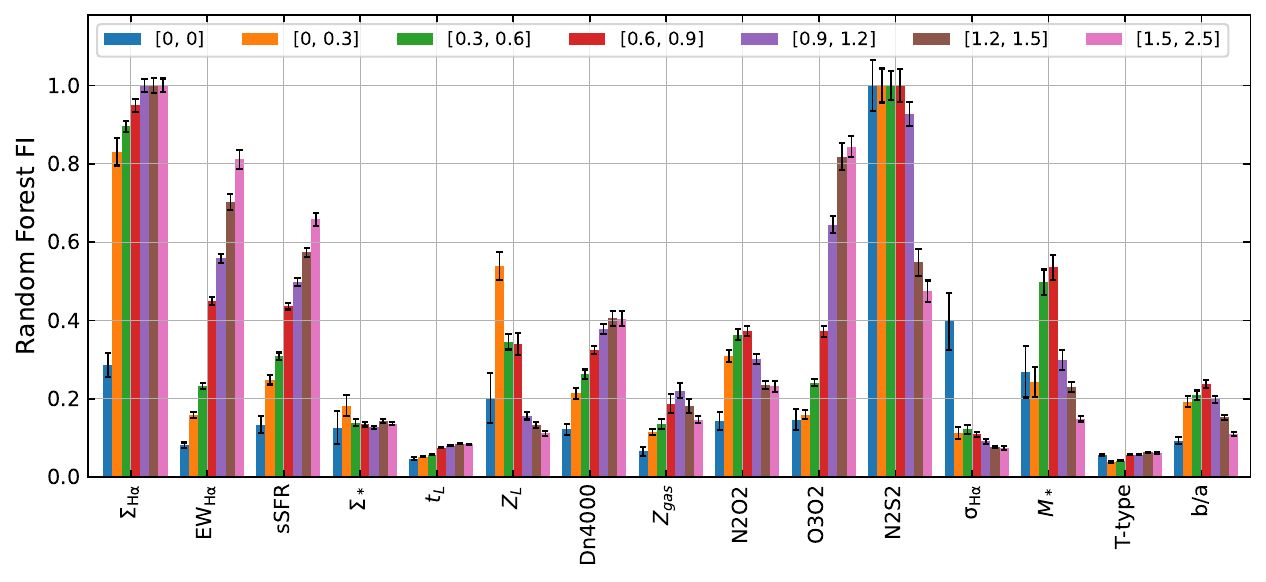}
	\caption{Random Forest regression is utilized to analyze the relationship between\
		\ebvgas\ and galactic properties across various radial bins,
		as indicated in the legend, using the unit of $R_e$. The histograms depict the 
		FI of the properties. It is noteworthy that the maximum FI in each radial bins
		has been normalized to 1.
	}
	\label{fig:lgbfi_r}
\end{figure*}

\section{Results} \label{sec:result}

We present our results in this section. In \autoref{sec:center},
we first focus on the gas-phase attenuation in ionized gas regions that are 
located less than 3\arcsec\ from the center of their host galaxy, for comparison 
with previous SDSS-based studies \citep[e.g.,][]{Maheson_G_2023}.
In \autoref{sec:allradii}, we extend the analysis of gas attenuation to the outer 
regions of our galaxies. Next, we consider both gas and stellar attenuations 
in \autoref{sec:radialprofiles}, and compare
H$\alpha$-bright and H$\alpha$-faint regions in \autoref{dig}. 
We note that, although we focus on gas attenuation in \autoref{sec:center} and \autoref{sec:allradii},
the same analysis is also done for stellar attenuation. The results are consistent 
with those presented in \autoref{sec:radialprofiles} and so are not included for simplicity.

\subsection{Feature importance to gas attenuation at galactic center}
\label{sec:center}
 
We first focus on the central regions of our galaxies.
We utilize the technique of Random Forest regression to analyze 
the relationship between \ebvgas\ and the 15 regional/global properties. 
In practice, we use the Random Forest regressor from the Python 
\texttt{Scikit-learn} package \citep{Pedregosa_F_2011}.
\autoref{fig:rf_center} displays the result of the feature importance (FI) 
analysis as the solid/blue histograms. In Panel (a),  following 
the approach outlined by \citet{Maheson_G_2023}, we exclude \hasb, 
sSFR and \haew\ from the analysis considering that all the three parameters 
are based on the H$\alpha$ flux which is corrected for dust attenuation 
through the Balmer decrement. Additionally, for this panel we omit \niisii, 
a parameter that was not considered in \citet{Maheson_G_2023}.
As can be seen, \havd\ and $M_\ast$ emerge as the most influential
properties in predicting \ebvgas, with subsequent importance
attributed to $Z_L$ and \oiiioii, both intricately linked to
stellar and gas metallicity. The results are in good agreement with
\citet{Maheson_G_2023}, demonstrating that the scaling relations
derived from single-fiber spectroscopy data within the SDSS are
replicable when our analysis is confined to the central region of
MaNGA galaxies. 

Next, we consider all the 15 properties. As shown in Panel (b) of 
\autoref{fig:rf_center}, \niisii\ emerges as the most important property, 
with a feature importance that is much higher than that of any other 
property. The most important property in Panel (a), \havd\ is now 
ranked in the second position, followed by $M_\ast$ and \hasb\ 
which exhibit comparable levels of significance. Previous studies 
have shown that \niisii\ is sensitive to both metallicity and ionization 
parameters. Specifically, this parameter exhibits an increasing 
trend toward higher metallicity (ionization parameter) when 
the ionization parameter (metallicity) is fixed \citep{Dopita_MA_2013}. 
As mentioned, \niisii\ was found to strongly correlate with \ebvgas\ 
at kpc scales in MaNGA galaxies regardless 
of radial distance \citep{Lin_ZS_2020,Li_N_2021}.

The smearing effect duo to the limited spatial resolution 
	of both SDSS and MaNGA may lead to underestimation of the parameters 
	in galactic centers. To test the potential effect of smearing on the FI analysis,
	we select all the spaxels from those galaxies at $z<0.03$, where the 
	physical resolution is better than the full sample.  The results of the FI 
	analysis for this subsample are shown in \autoref{fig:rf_center} as red histograms. 
	For the analysis with the subset of parameters as shown in the left panel, 
	obvious variation is seen between the full sample and the subsample, 
	particularly for $M_\ast$, $\Sigma_\ast$ and $Z_L$. At $z<0.03$ where 
	the smearing effect is less severe, $M_\ast$ is no longer the most important 
	parameter, while regional properties $\Sigma_\ast$ and $Z_L$ become 
	more important. In contrast, for the FI analysis with all the 15 parameters
	as shown in the right panel, the results of the subsample are consistent with 
	that of the full sample. This test suggests that the FI analysis is robust to 
	the smearing effect as long as the truly relevant parameters are included.
	It is interesting to note that the FI of $\sigma_{\text{H}\alpha}$ shows 
	little variation between the two samples, a result that is true in both panels.

\begin{figure*}
	\fig{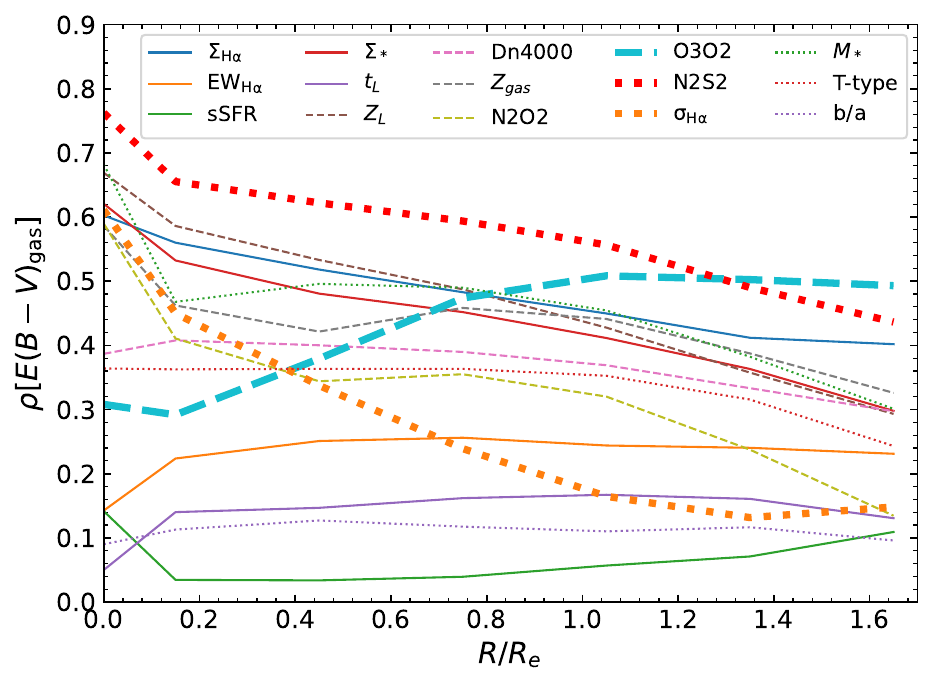}{0.4\textwidth}{(a)}
	\fig{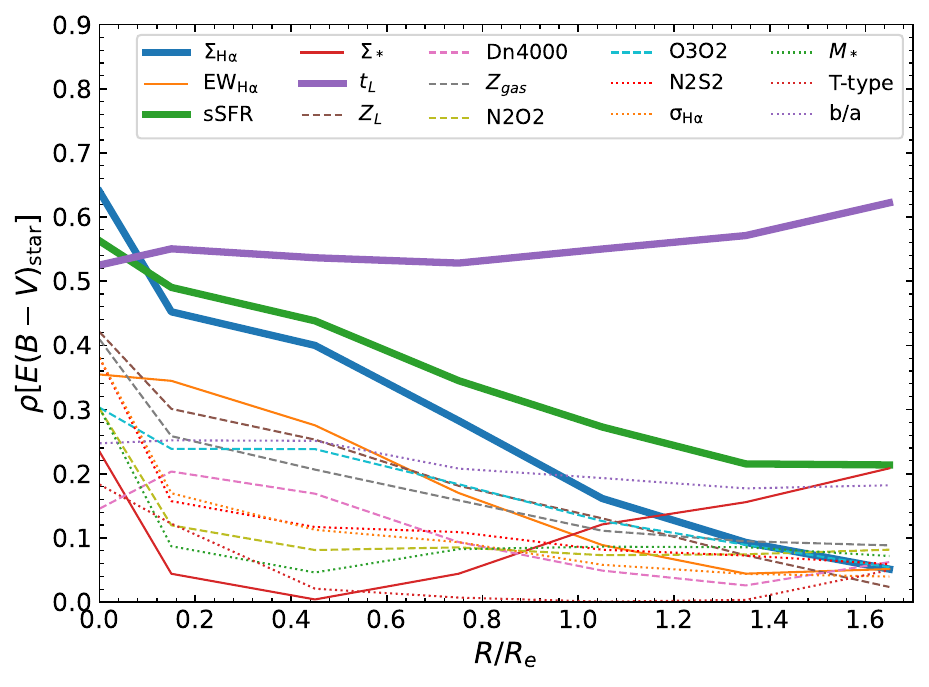}{0.4\textwidth}{(b)}
	\caption{Panel (a): Radial profiles of correlations between \ebvgas\ and
		galactic properties.
		The x-axis represents the radial distance scaled by $R_e$ from the
		galactic center, while the y-axis depicts the Spearman rank correlation
		coefficients between \ebvgas\ and various galactic properties at different
		radii. Panel (b): The symbols and lines are the same as in Panel (a)
		but for \ebvstar.
	}
	\label{fig:radial_profile_0}
\end{figure*}

\subsection{Feature importance to gas attenuation at all radii}
\label{sec:allradii}

We then consider all the regions and examine how the feature importance 
to \ebvgas\ may vary as one goes from the galactic center to larger radial 
distances. 
In \autoref{fig:lgbfi_r}, we present the FI analysis depicting the
relationship between the 15 galactic properties and \ebvgas\ across
various radial regions from the center to the outer regions.
Different colors correspond to different radial bins as indicated.
With an increase in distance from the galactic center, several noticeable
results can be read from the figure. First, \niisii\ consistently maintains 
a high FI across almost all the radii. Specifically, \niisii\ is the leading 
parameter out to $\sim R_e$, before it gives place to \hasb\ at 
$0.9R_e<R<1.2R_e$ with a FI that is only slightly smaller than that of 
the new leader. Although its FI further drops at larger radii, \niisii\ ends up 
with a relatively high FI, taking the fifth position in the largest radial bin. 
Second, the most important parameters identified previously in 
\citet{Maheson_G_2023}, the FI of $M_\ast$, $Z_L$ and \havd\ decreases 
with increasing distance. In particular, 
\havd\ experiences an immediate and dramatic decline in FI 
as one goes beyond the first radial bin `[0, 0]', strongly indicating that
this parameter is only important in the galactic center. Third, properties 
associated with H$\alpha$ flux such as \hasb, \haew\ and sSFR 
exhibit much higher FI in the outer regions than in the galactic center, 
and they become the dominate parameters in the largest radial bin. 
Finally, properties linked to gas metallicity and ionization, \niioii\ and 
\oiiioii\ manifest significantly greater FI at larger radii. In the outermost 
regions, \oiiioii\ even presents a higher FI than \niisii.

Apparently, our findings indicate a pronounced radial variability
in the FI of galactic properties to \ebvgas. This variability introduces a 
heightened level of complexity to the gas attenuation scaling relations,
when compared to previous studies which have been mostly limited 
to either the galactic center or the whole galaxy.

\begin{figure*}
	\epsscale{1.15}
	\plotone{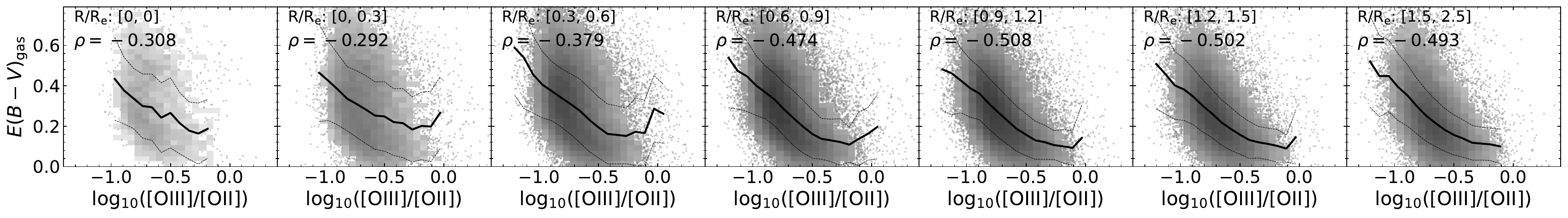}
	\par
	\plotone{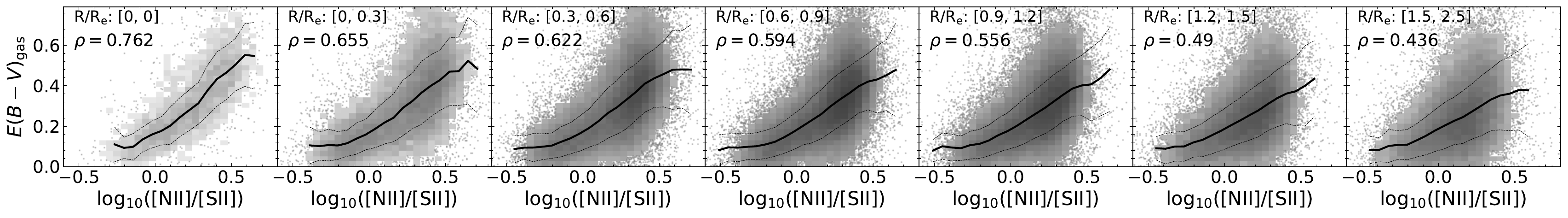}
	\par
	\plotone{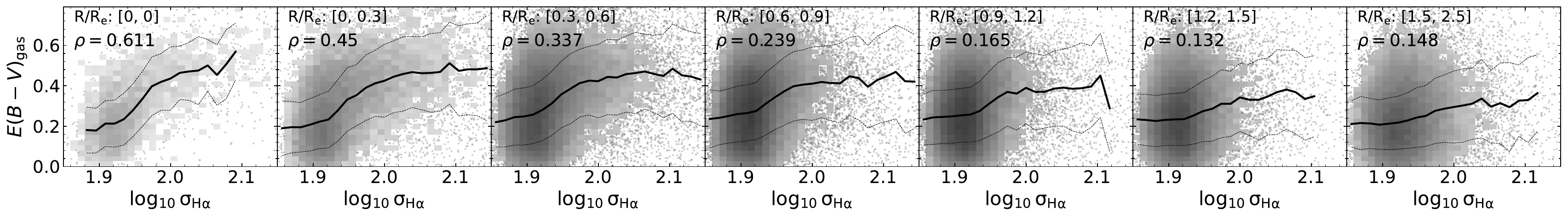}
	\caption{The three rows describe correlations between \ebvgas\ and \oiiioii, \niisii, \havd. 
		The gray-scale background in each panel represents the distributions of all
		ionized gas regions, with the median relation and 1$\sigma$ scatter delineated
		by black lines. The Spearman rank correlation coefficients ($\rho$) 
		and the range of distance to the galactic center scaled in $R_e$
		are indicated in the upper-left corner.
	}
	\label{fig:corr_ebv_0}
\end{figure*}

\begin{figure*}
	\epsscale{1.15}
	\plotone{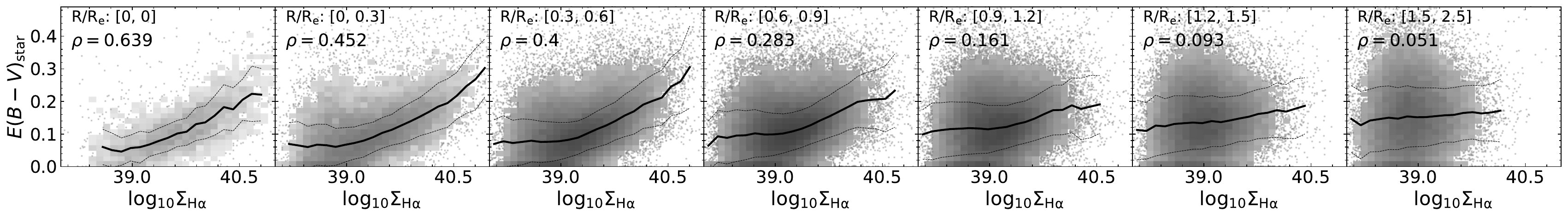}
	\par
	\plotone{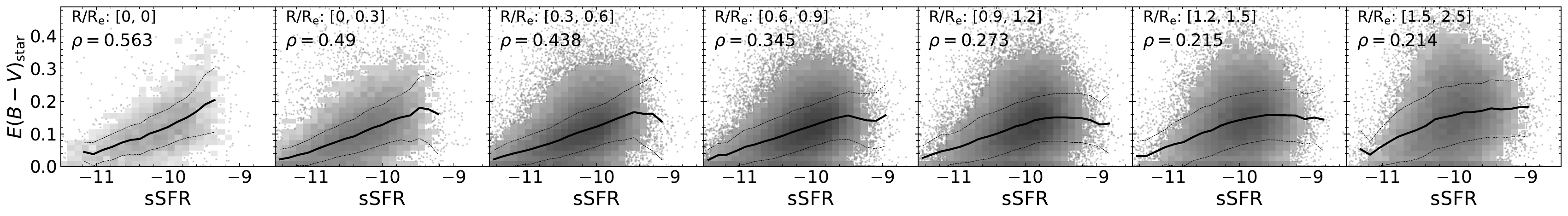}
	\par
	\plotone{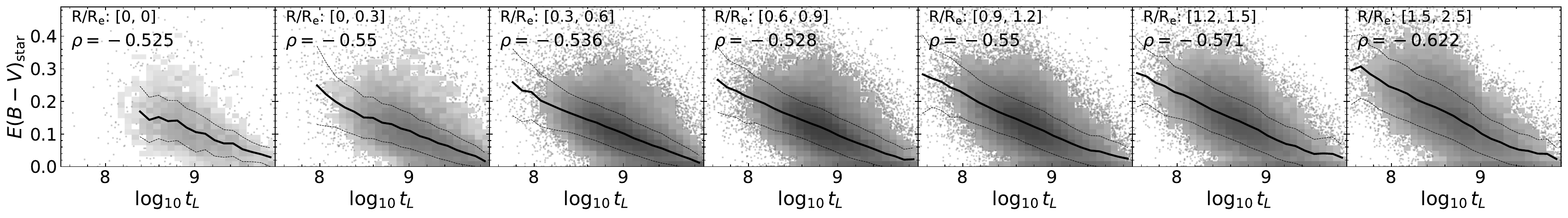}
	\caption{The three rows depict correlations between \ebvstar\ and \hasb, sSFR, $t_L$.
		The symbols and lines are the same as in \autoref{fig:corr_ebv_0}.
	}
	\label{fig:corr_ebv_1}
\end{figure*}

\subsection{Radial profiles of dust attenuation scaling relation}
\label{sec:radialprofiles}

To enable a direct and more intuitive comparison of the radial variations of 
the dust attenuation scaling relations, we further utilize the Spearman rank 
correlation coefficients ($\rho$) to quantify the correlation strength 
of both \ebvgas\ and \ebvstar\ with all the 15 properties and in all the radial bins. 
The result of this analysis is shown in \autoref{fig:radial_profile_0}, 
where the two panels show the radial profiles of $\rho$ for \ebvgas\ 
and \ebvstar, respectively. In each panel the different lines/colors 
correspond to different properties, as indicated. Note that, we show 
the {\em absolute} values of $\rho$, thus comparing the strength of 
correlations while ignoring whether the correlations are positive or negative.

For \ebvgas, as seen from Panel (a) of \autoref{fig:radial_profile_0}, 
both the relative ranks of the correlation coefficient for different properties 
at given radial distance and the radial variations of the correlation coefficient 
for given property are generally consistent with what we have seen 
above from the analysis of FI provided by Random Forest. For instance, 
\niisii\ presents the highest value of $\rho$ out to $R\sim1.3R_e$, 
ranging from $\rho=0.76$ at $R=0$ to $\rho\sim0.5$ at $1.3R_e$, 
while \havd\ shows the maximum decline, starting with a relatively high 
value of $\rho\sim0.6$ and ending with $\rho\sim0.15$ at the largest radius. 
In fact, a negative gradient is seen in all the eight properties that exhibit 
relatively strong correlations ($\rho>0.5$) with \ebvgas\ at the galactic 
center. Ordered by decreasing $\rho$ at $R=0$, these properties are  
\niisii, $M_\ast$, $Z_L$, \sigmamass,  \havd, \hasb, 12+\logten(O/H), 
and  \niioii. In contrast, all the other 7 properties except \oiiioii\ 
show no/weak radial gradient, thus maintaining relatively weak correlations 
($\rho<0.5$) at all radial distances. The only exception, \oiiioii\
presents significant increase with radius, starting with a low value of 
$\rho\sim0.3$ at $R=0$ and reaching the maximum value of $\rho\sim0.5$ 
at $R>1.3R_e$, even surpassing \niisii. 
\autoref{fig:corr_ebv_0} 
displays the distribution of individual regions on the diagrams formed 
by \ebvgas\ against \oiiioii, \niisii\ and \havd. Panels from left to right 
correspond to different distances from the galactic center, as indicated. 
The median relation and the 1$\sigma$ scatter are plotted as the thick 
line and the two thin lines in each panel. We see that, \oiiioii\ exhibits 
an increasingly strong and negative correlation with \ebvgas\ as the 
radius increases, while \niisii\ presents a strong correlation with \ebvgas\ 
in the galactic center and maintains a relatively strong correlation in the
outer regions. Regarding \havd, a strong correlation is found only in 
the first radial bin, with $\rho$ declining rapidly in the outer regions.

For \ebvstar, as seen from Panel (b) of \autoref{fig:radial_profile_0}, 
the correlation coefficient declines for all properties except $t_L$ 
as one goes beyond the galactic center. As a result, in the outermost 
region $t_L$ becomes the only dominant parameter while all the other 
properties are weakly correlated with \ebvstar. In the galactic center, 
only three properties show relatively strong correlations with $\rho>0.5$: 
\hasb, sSFR, and $t_L$. Although \hasb\ and sSFR exhibit a stronger
correlation with \ebvstar\ than $t_L$ in the center, their correlations 
rapidly decrease with increasing radius. The correlation coefficient of 
$t_L$ is roughly constant at $\rho\sim0.55-0.6$ at all radii. 
These results can also be seen from \autoref{fig:corr_ebv_1} which 
displays the individual regions on the diagrams of \ebvstar\ 
versus \hasb, sSFR and $t_L$ in different radial bins. 

In conclusion, \niisii\ is the main driver of \ebvgas, while $t_L$
drives \ebvstar\ from the center to the outer regions, consistent with
our findings in \citetalias{Li_N_2021}. Several properties exhibit a
relatively strong correlation with \ebvgas\ in the center, although
these correlations notably weaken as the radius increases.
Regarding \ebvstar, only SFR-related properties are comparable to
correlation of $t_L$ in the central regions.

\begin{figure*}
    \fig{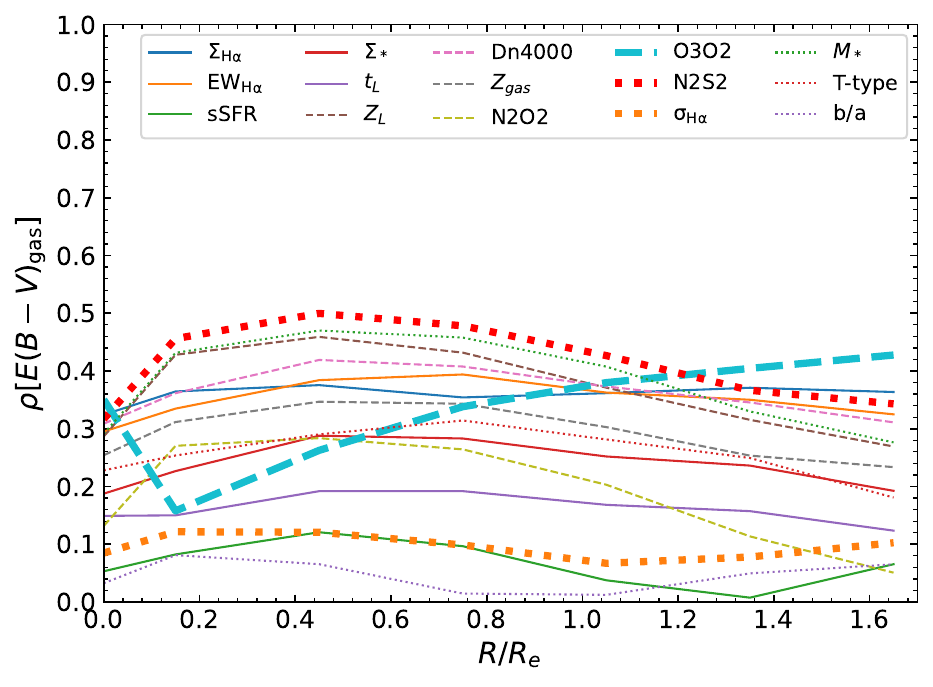}{0.4\textwidth}{(a) \hasb$\leq10^{39}$\hasbunit}
    \fig{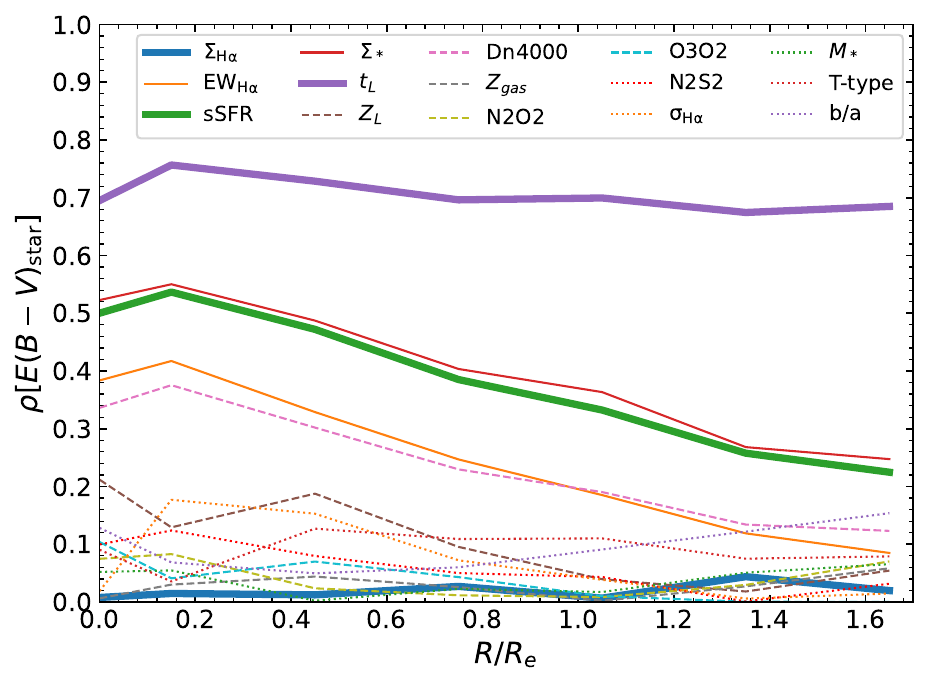}{0.4\textwidth}{(b) \hasb$\leq10^{39}$\hasbunit}
    \par
    \fig{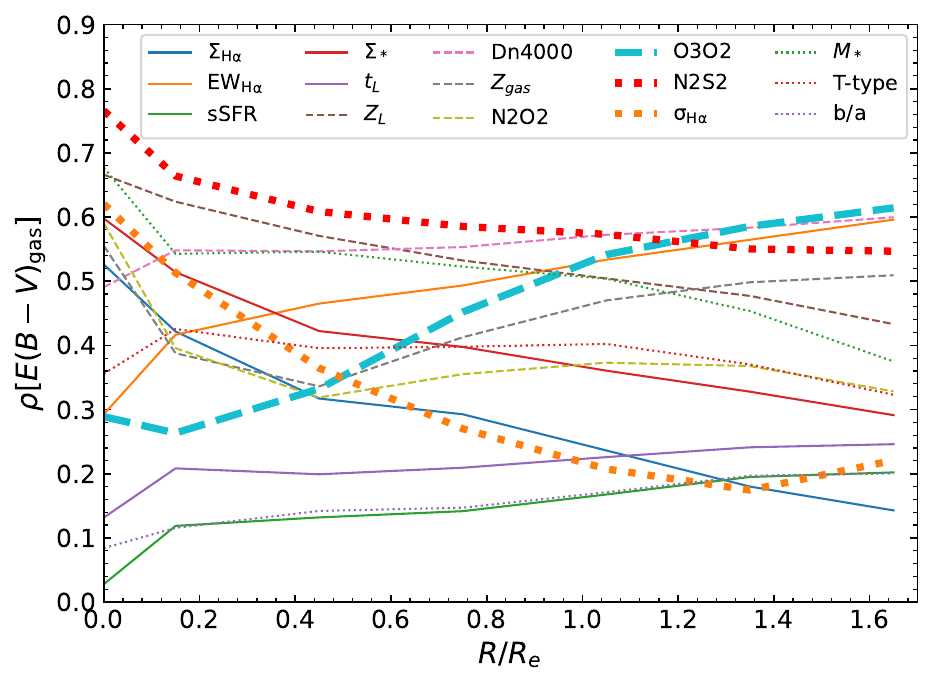}{0.4\textwidth}{(c) \hasb$>10^{39}$\hasbunit}
    \fig{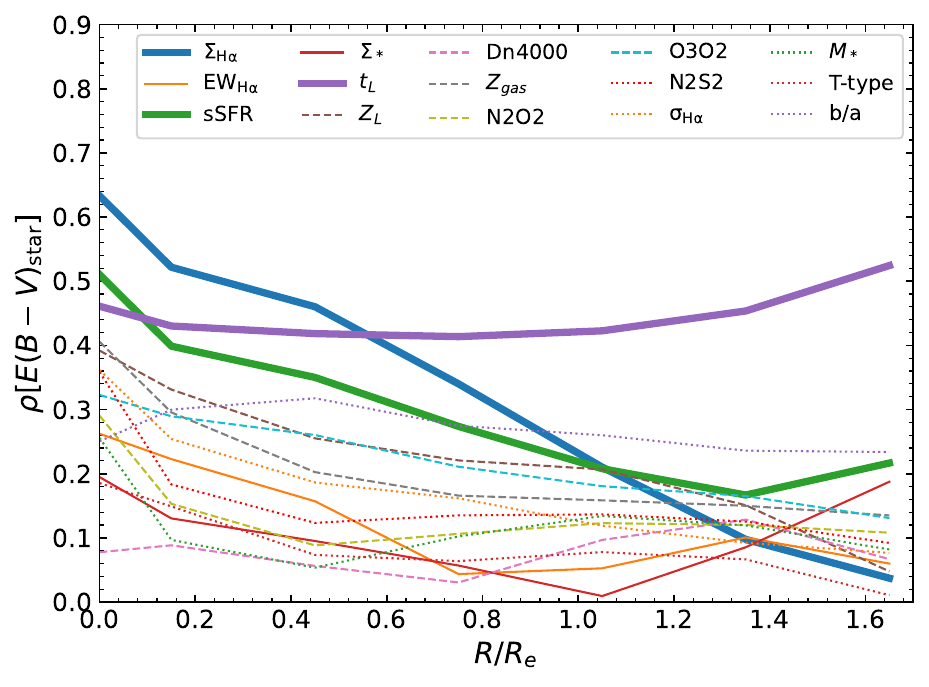}{0.4\textwidth}{(d) \hasb$>10^{39}$\hasbunit}
    \caption{Panel (a) and Panel (c) are Radial profiles of correlations
    between \ebvgas\ and galactic properties in H$\alpha$-faint regions with \hasb$\leq10^{39}$\hasbunit\
    and H$\alpha$-bright regions with \hasb$>10^{39}$\hasbunit, respectively.
    Panel (b) and Panel (d) are Radial profiles of correlations
    between \ebvstar\ and galactic properties in H$\alpha$-faint and H$\alpha$-bright regions, respectively.
    The symbols and lines are the same as in \autoref{fig:radial_profile_0}.
    }
    \label{fig:radial_profile_1}
\end{figure*}

\subsection{Radial profiles of dust attenuation scaling relation in H$\alpha$-faint and H$\alpha$-bright regions}
\label{dig}

\autoref{fig:radial_profile_1} displays the radial profiles of the correlation 
coefficient $\rho$ for \ebvgas\ (left panels) and \ebvstar\ (right panels),
and for H$\alpha$-faint (upper panels) and H$\alpha$-bright (lower panels) regions separately. 
For \ebvgas, we see that all of the properties present moderate or weak 
correlations ($\rho<0.5$) in H$\alpha$-faint regions, while the H$\alpha$-bright regions show 
similar results to what we have seen in Panel (a) of the previous figure. 
This indicates that the scaling relations of \ebvgas\ observed for the 
full sample are predominantly contributed by H$\alpha$-bright regions. In addition, 
comparing the H$\alpha$-bright regions with the full sample, we find \haew, $D_n4000$, 
and $Z_L$ to display notably stronger correlations with \ebvgas, 
particularly in outer regions with $R\ga R_e$. Even in the outermost 
regions, five properties still maintain $\rho>0.5$, with three of them 
displaying stronger correlations with \ebvgas\ compared to \niisii. 
This complexity adds challenges to determining the true driving factor 
among these properties or establishing a scaling relation for estimating 
\ebvgas.

For \ebvstar, the result appears less complex in the sense that 
only a few properties show relatively strong correlations with $\rho>0.5$
in both H$\alpha$-faint and H$\alpha$-bright regions and at any given radius. Nonetheless, 
the two types of regions are different in several aspects. 
First, if we only consider H$\alpha$-faint regions, we find the strong correlation 
of \ebvstar\ with $t_L$ as seen from Panel (b) of the previous figure 
remains similarly strong or even stronger over the full range of radial 
distance, with $\rho\sim0.7$ here versus $\rho\sim0.6$ in \autoref{fig:radial_profile_0}. 
In contrast, for H$\alpha$-bright regions the correlation of \ebvstar\ with $t_L$ is 
only moderate, with $\rho\sim0.5$, though still independent of radial 
distance. Second, \hasb\ in H$\alpha$-faint regions shows no correlation with 
\ebvstar\ at all radii, while \hasb\ in H$\alpha$-bright regions presents a radial profile 
of $\rho$ similar to that of the full sample. Third, different 
from both $t_L$ and \hasb, sSFR presents similar profiles in H$\alpha$-faint and 
H$\alpha$-bright regions, with a moderate coefficient of $\rho\sim0.5$ at $R=0$ 
and a low coefficient of $\rho\sim0.2-0.3$ in the outermost region. 
Finally, we notice that, although \sigmamass\ in H$\alpha$-bright regions 
roughly follows the radial profile of the full sample, with rather weak 
correlations ($\rho\lesssim0.2$) at all radii, this parameter presents
almost the same profile as sSFR in H$\alpha$-faint regions. This result may 
be understood from the fact that \sigmamass\ is largely contributed 
by old stellar populations which are mostly associated with H$\alpha$-faint 
regions. Considering both that H$\alpha$-faint and H$\alpha$-bright regions are dominated 
respectively by \sigmamass\ and \hasb\ and that sSFR is defined by 
the ratio of SFR (which can be estimated from \hasb) to \sigmamass, 
it is natural to see that sSFR behaves similarly to \sigmamass\ in H$\alpha$-faint 
regions but similarly to \hasb\ in H$\alpha$-bright regions. 

\begin{figure*}
	\epsscale{1.15}
	\plotone{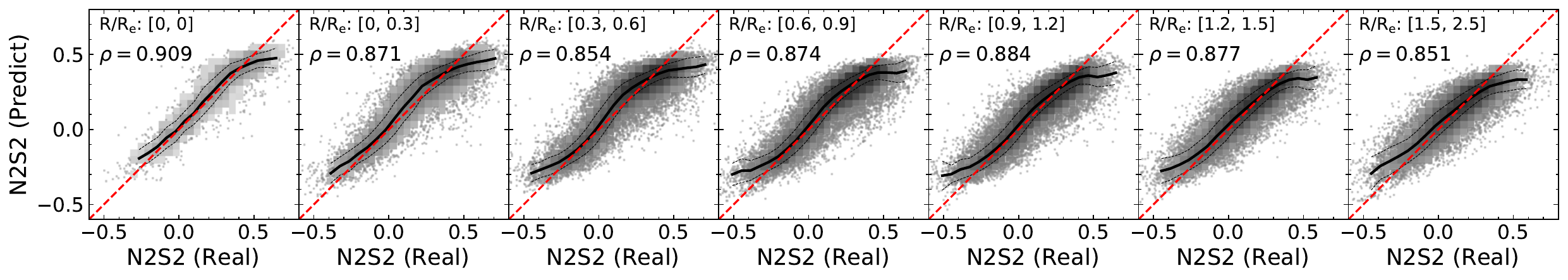}
	\plotone{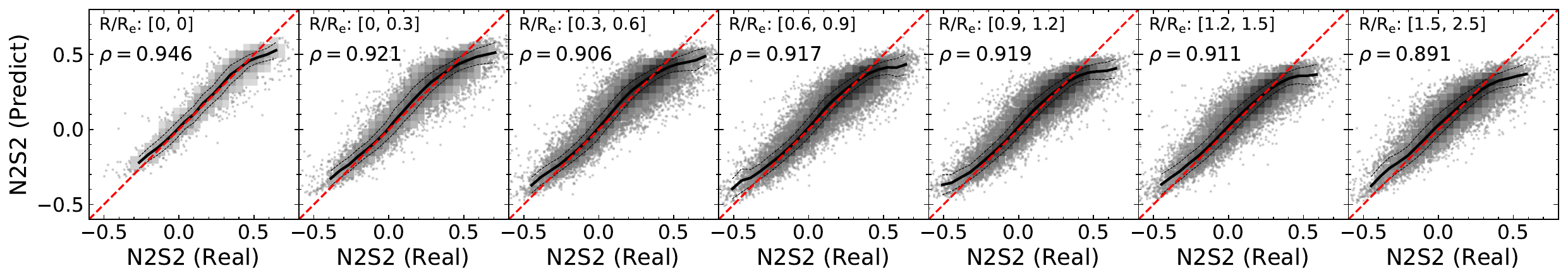}	
	\caption{Predicted \niisii\ versus the observed value for ionized gas regions 
		in different radial bins as indicated in each panel. In the upper panels, 
		the prediction is made by using \sigmamass\ and 12+\logten(O/H) as inputs 
		to train a Random Forest regressor. In the lower panels, \oiiioii\ is used in 
		addition to \sigmamass\ and 12+\logten(O/H) for the prediction. 
		In each panel, the dashed red line represents the 1:1 relation, while 
		the solid thick and thin lines indicate the median and the 1$\sigma$ 
		scatter of the relation.}
	\label{fig:n2s2_predict}
\end{figure*}

\section{Discussion} \label{sec:discussion}

\subsection{Gas attenuation}

Our work has clearly established that \niisii\ outperforms all the other 
properties as the most important parameter in relation to \ebvgas. 
This result was found in \citetalias{Li_N_2021} for all the ionized gas regions 
as a whole. We further found here that the conclusion is true for a wide 
range of radial distance from the galactic center out to $R\ga R_e$, 
and it holds for both H$\alpha$-faint and H$\alpha$-bright regions although the overall correlations 
in H$\alpha$-faint regions are relatively weak. The physical reason behind the high 
importance of \niisii\ to gas attenuation is not immediately clear, however. 
When \niisii\ is excluded from the analysis, both stellar mass and metallicity 
are highly ranked in feature importance as shown in \autoref{fig:rf_center}, 
consistent with the previous SDSS-based study by \citet{Maheson_G_2023}. 
As pointed out by \citet{Maheson_G_2023}, the dependence 
of dust attenuation on stellar mass and metallicity is expected from
simple analytical equations relating dust mass with gas mass and metallicity,
as well as the molecular gas main sequence for resolved regions in 
ALMaQUEST \citep[MGMS][]{Lin_L_2019}. Since \niisii\ is commonly employed 
as a diagnostic for gas-phase metallicity
\citep[e.g.,][]{Perez-Montero_E_2009,Dopita_MA_2016,Teklu_BB_2020}
and given the known stellar mass-metallicity relation
\citep[e.g.,][]{Tremonti_CA_2004,Koppen_J_2007,Yates_RM_2012,Dayal_P_2013,Ma_XC_2016},
one can well expect \niisii\ to also strongly depend on \ebvgas. 
It is likely that \niisii\ integrates the favorable factors in both stellar mass and 
metallicity, thus rendering it the most sensitive property to variations in \ebvgas, 
and this is the reason why the feature importance of both stellar mass and metallicity 
is reduced dramatically when \niisii\ is included in the Random Forest analysis. 
To test this conjecture, we train a Random Forest regressor using \sigmamass\ 
and 12+\logten(O/H) as inputs to predict \niisii. In the upper panels of \autoref{fig:n2s2_predict}, 
we compare the {\em predicted} \niisii\ with the observed value for ionized gas 
regions in different radial bins, with the correlation coefficient $\rho$ indicated 
in each panel. As expected, \niisii\ can be reasonably well predicted by the 
two properties, with $\rho> 0.85$ in all cases. This prediction is most accurate 
in the galactic center ($\rho=0.91$), and becomes slightly less accurate as 
the radial distance increases. 

Another important parameter identified in our work is \oiiioii\ which 
presents similarly strong or even stronger correlations with \ebvgas\ 
in the outer regions, when compared to \niisii. In fact, \oiiioii\ as an 
indicator of ionization parameter and nitrogen enrichment can also 
influence \niisii. At a fixed metallicity, \niisii\ can increase with ionization 
parameter or primary nitrogen enrichment \citep{Dopita_MA_2013,Blanc_GA_2015}. 
We have repeated the above analysis, using \oiiioii\ in addition to 
\sigmamass\ and 12+\logten(O/H) as inputs to train the Random Forest regressor. 
The result is shown in the lower panels of \autoref{fig:n2s2_predict}. 
We see that the correlation between the predicted and real \niisii\ 
becomes slightly tighter at given radial bin, with an average increase 
of $\sim5\%$ in the correlation coefficients. On one hand, this result 
confirms that \niisii\ and \oiiioii\ indeed share common information 
(e.g. ionization parameter) that is closely related to dust attenuation in 
ionized gas. On the other hand, however, the limited improvement 
in predicting \niisii\ by the inclusion of \oiiioii\ strongly implies that 
\oiiioii\ represents some important and distinct factor that is not fully 
encoded in \niisii, and this factor is very likely the ionization parameter, 
which plays a more important role in gas attenuation in the outer regions 
of galaxies when compared to \niisii. 

The nebular velocity dispersion as quantified by \havd\ was previously 
found to be important in determining gas attenuation based on 
SDSS single fiber spectra \citep{Maheson_G_2023}. Those authors 
suggested that the unexpected importance of \havd\ may be understood 
if the nebular velocity dispersion trances the gravitational potential in 
the galaxy, which determines how much the galaxy can retain dust 
and metals against radiation pressure and gas outflows \citep{Chisholm2015}.
Our finding that the strong correlation of \havd\ with gas attenuation 
holds only in the galactic centers provides further evidence in support 
of their conjecture, considering that the gravitational potential well is 
deeper in the center than the outer regions. 
Alternatively, \havd\ may be tracing non-gravitational motions 
	such as outflows, which may also play important roles by pushing the dust 
	to large radii, thus reducing the content of dust in the central region.	
	In addition, the importance of \havd\  for gas attenuation
	could be related to the contamination of low-ionization nuclear emission 
	regions (LINERs) in galactic centers. As found in \citet{Law_DR_2021},  
	regions typically associated with AGN or LINER exhibit much higher 
	\havd\ values (reaching 100-200 km/s) when compared to \hii\ regions located 
	at larger radii. \autoref{fig:corr_ebv_0} suggests that the 
strong correlation between \havd\ and \ebvgas\ in galactic centers 
is largely attributed to the component with \havd\ greater than 100 km/s.
More work would be needed in future if one were to fully understand the 
important role of \havd\ for gas attenuation in galactic centers.

\subsection{Stellar attenuation}

The strong correlation between \ebvstar\ and stellar age in resolved 
regions has been found in \citetalias{Li_N_2021} for all the ionized gas 
regions as a whole. The age-dependent stellar attenuation has also 
been considered in some earlier studies \citep[e.g.,][]{Panuzzo_P_2007,Noll_S_2009,
Buat_V_2012,Lo-Faro_B_2017,Tress_M_2019}. 
The result in our work further shows that the driving role of stellar age 
for \ebvstar\ holds across the whole galaxy, independent of radial distance, 
and the strong correlation is predominantly contributed by H$\alpha$-faint regions. 
This is expected considering that the H$\alpha$-faint regions are dominated by old 
stellar populations. Moreover, the importance of \hasb\ and sSFR in 
determining \ebvstar\ in the galactic center was overlooked by 
\citetalias{Li_N_2021} due to the contamination of outer regions 
where the importance of \hasb\ and sSFR is low. When dividing the 
ionized regions into H$\alpha$-faint and H$\alpha$-bright regions, we further find that the 
important role of \hasb\ and sSFR holds only in H$\alpha$-bright regions at the 
galactic center. This may be understood from the fact that the central 
star formation is usually associated with environments with increased 
dust content which in turn plays a vital role in the star formation process
\citep[e.g.,][]{Dwek_E_1998,Panuzzo_P_2007,Dwek_E_2011,Zhukovska_S_2014}.

\subsection{Implications for high-$z$ studies}

The strong dependence of dust attenuation on stellar mass has 
been found also for galaxies at higher redshifts, and recent studies 
have consistently reported a lack of significant evolution in the 
relation between dust attenuation and stellar mass over a wide 
range of redshift from $z\sim0$ up to $z\sim6.5$
\citep[e.g.][]{Shapley_AE_2022,Shapley_AE_2023,Maheson_G_2023}. 
Studies of high$-z$ galaxies are mostly limited to either measurements 
of global properties due to poor angular resolution of ground-based 
observations, or small samples of spatially resolved properties from 
space telescopes with poor spectral resolution. In most cases 
where ground-based observations are used, the best available seeing 
$\sim0.6$\arcsec\ roughly corresponds to $\sim4-5$ kpc at $z>1$. 
Our work has revealed strong radial variations in the dust attenuation 
scaling relations in nearby galaxies, which (if hold true also at higher 
redshifts) could be barely resolved with current observations of 
high-$z$ galaxies. It is unclear to what extent the lack of evolution 
in the attenuation-mass relation found in previous studies is caused 
by the poor angular resolution of the data. In addition, the significantly
different attenuation relations as found in H$\alpha$-faint and H$\alpha$-bright regions 
need to be considered when interpreting results at high redshifts,
because the relative contribution of star-forming regions to the 
global dust attenuation relations is expected to be a function of 
redshift given the evolution of the cosmic star formation rate density
\citep[e.g.][]{Madau2014}.

\section{Summary} \label{sec:summary}

This is the third paper of a series of studies on resolved dust attenuation 
in nearby galaxies. In the previous two papers, we have developed a new
technique to estimate a model-independent attenuation curve from a given
optical spectrum of MaNGA galaxies \citepalias{Li_N_2020},
which has enabled us to study the correlations of both stellar and gas 
attenuations at kpc scales with a large number of regional/global properties 
of nearby galaxies \citepalias{Li_N_2021}. In this work we have extended 
our study by further examining the radial variations of the dust attenuation 
relations, using the same data as in \citetalias{Li_N_2021}. We apply 
the Random Forest regression technique to obtain the feature importance 
(FI) of all the regional/global properties in relation to \ebvgas\ and \ebvgas.
We also calculate the Spearman correlation coefficients ($\rho$) to 
quantify the strength of the correlations between dust attenuation and 
galactic properties. We perform these analyses for ionized regions in 
different radial bins separately, and for H$\alpha$-bright regions and 
H$\alpha$-faint regions separately.

First of all, the dust attenuation scaling relations obtained 
previously from single-fiber spectroscopy data within the SDSS are 
reproducible when our analysis is confined to {\em the central region} 
of MaNGA galaxies.
\begin{itemize}
\item  If \niisii\ is not included in the analysis as in previous studies, 
nebular velocity dispersion (\havd), global stellar mass ($M_\ast$), 
luminosity-weighted stellar metallicity ($Z_L$) and surface stellar 
mass density (\sigmamass) are found to be most important for 
\ebvgas. This result is in good agreement with previous SDSS-based studies. 
If included, \niisii\ outperforms all other properties as the most important 
property in the galactic center in relation to \ebvgas. 

\item For \ebvgas, following \niisii, quite a number of properties including 
\havd, $M_\ast$, \sigmamass, $Z_L$, H$\alpha$ surface brightness (\hasb), 
gas-phase metallicity (12+\logten O/H) and the \niioii\ line ratio (N2O2) 
exhibit relatively high feature importance and strong correlation coefficient. 
For \ebvstar, in contrast, only a few properties (\hasb, sSFR, and $t_L$) 
show relatively strong correlations in the galactic center.
\end{itemize}  

When considering regions at all radii, the scaling relations of both 
\ebvgas\ and \ebvstar\ are found to strongly vary as one goes from 
the galactic center towards outer regions, and from H$\alpha$-faint regions and
H$\alpha$-bright regions. 
\begin{itemize}
	\item For \ebvgas, \niisii\ is top ranked in feature importance and 
	presents a much higher correlation coefficient than any other properties 
	and over a wide range of radial distance from $R=0$ out to $R\sim R_e$.  
	In the outermost regions ($R>1.2R_e$), \oiiioii\ outperforms \niisii\ 
	as the leading property in relation to \ebvgas, although \niisii\ still 
	remains a comparably high correlation coefficient.  
   \item For \ebvstar, stellar age $t_L$ shows a strong correlation
    with $\rho\sim0.55-0.6$ only weakly dependent on radial distance, 
    making itself the most important property over all radii except 
    the galactic center where H$\alpha$ surface density (\hasb) 
    and specific star formation rate (sSFR) present similarly strong 
    correlations. The correlation strengths of the latter two properties 
    decline rapidly with radial distance. 
    \item When dividing the ionized regions into H$\alpha$-bright regions and H$\alpha$-faint regions, 
    we find the former to generally show stronger correlations with \ebvgas\ 
    and the latter to be more strongly correlated with \ebvstar, although 
    depending on individual properties and radial distance.
\end{itemize}

 \section*{Acknowledgements}

We are grateful to the anonymous referee whose comments 
have helped improve this paper.
This work is supported by the National Key R\&D Program of China (grant NO. 2022YFA1602902), 
and the National Natural Science Foundation of China (grant Nos. 12433003, 11821303, 11973030).

Funding for the Sloan Digital Sky Survey IV has been provided by the
Alfred P. Sloan Foundation, the U.S. Department of Energy Office of
Science, and the Participating Institutions. SDSS-IV acknowledges
support and resources from the Center for High-Performance Computing
at the University of Utah. The SDSS web site is www.sdss.org.

SDSS-IV is managed by the Astrophysical Research Consortium for the
Participating Institutions of the SDSS Collaboration including the
Brazilian Participation Group, the Carnegie Institution for Science,
Carnegie Mellon University, the Chilean Participation Group, the
French Participation Group, Harvard-Smithsonian Center for
Astrophysics, Instituto de Astrof\'isica de Canarias, The Johns
Hopkins University, Kavli Institute for the Physics and Mathematics of
the Universe (IPMU) / University of Tokyo, Lawrence Berkeley National
Laboratory, Leibniz Institut f\"ur Astrophysik Potsdam (AIP),
Max-Planck-Institut f\"ur Astronomie (MPIA Heidelberg),
Max-Planck-Institut f\"ur Astrophysik (MPA Garching),
Max-Planck-Institut f\"ur Extraterrestrische Physik (MPE), National
Astronomical Observatories of China, New Mexico State University, New
York University, University of Notre Dame, Observat\'ario Nacional /
MCTI, The Ohio State University, Pennsylvania State University,
Shanghai Astronomical Observatory, United Kingdom Participation Group,
Universidad Nacional Aut\'onoma de M\'exico, University of Arizona,
University of Colorado Boulder, University of Oxford, University of
Portsmouth, University of Utah, University of Virginia, University of
Washington, University of Wisconsin, Vanderbilt University, and Yale
University.

%



\software{Astropy \citep{2013A&A...558A..33A,2018AJ....156..123A},
Scikit-learn \citep{Pedregosa_F_2011}}






\bibliography{ref}{}
\bibliographystyle{aasjournal}



\end{document}